\documentclass[sigconf]{acmart}

\usepackage[font=small,labelfont=bf]{caption}
\renewcommand\footnotetextcopyrightpermission[1]{} 
\acmConference{}
\acmYear{}
\usepackage{hyperref}

\usepackage{tikz}
\usepackage{amsmath}
\usepackage{filecontents}
\usepackage{footnote}
\usepackage{diagbox}
\usepackage{listings}
\usepackage{booktabs}
\usepackage{multirow}
\usepackage[ruled,vlined]{algorithm2e}
\usepackage{subfigure}
\usepackage{tabularx}
\usepackage{url}
\usepackage{amsfonts}
\usepackage{algorithmic}
\usepackage{graphicx}
\usepackage{textcomp}
\usepackage{xcolor}
\usepackage[skins]{tcolorbox}
\usepackage{verbatim}
 
\usepackage{enumitem}
\usepackage{pifont}
\usepackage{listings}
\usepackage[framemethod=tikz]{mdframed}
\usepackage{balance}
\usepackage{lipsum} 
\usepackage{indentfirst}

\def\eg{\emph{e.g.,}\xspace}

\def\etal{\emph{et al.,}\xspace}

\newcommand{\fancyref}[1]{\etal~\cite{#1}}
\newcommand{\one}{({\em i}\/)}
\newcommand{\two}{({\em ii}\/)}
\newcommand{\three}{({\em iii}\/)}
\newcommand{\four}{({\em iv}\/)}
\newcommand{\five}{({\em v}\/)}

\definecolor{dkgreen}{rgb}{0,0.6,0}
\definecolor{gray}{rgb}{0.5,0.5,0.5}
\definecolor{mauve}{rgb}{0.58,0,0.82}
\definecolor{applegreen}{rgb}{0.55, 0.71, 0.0}
\definecolor{amber}{rgb}{1.0, 0.75, 0.0}
\definecolor{firebrick}{rgb}{0.7, 0.13, 0.13}
\definecolor{darkblue}{rgb}{0,0,0.55}

\def\BibTeX{{\rm B\kern-.05em{\sc i\kern-.025em b}\kern-.08em
    T\kern-.1667em\lower.7ex\hbox{E}\kern-.125emX}}
\begin{document}

\title{Statically Detecting Adversarial Malware\\ through Randomised Chaining} 

\author{Matthew Crawford}
\affiliation{%
  \institution{University of Adelaide}
  \country{Australia}
}

\author{Wei Wang}
\affiliation{%
  \institution{University of Adelaide}
  \country{Australia}
}
\author{Ruoxi Sun}
\affiliation{%
  \institution{University of Adelaide}
  \country{Australia}
}

\author{Minhui Xue}
\affiliation{%
  \institution{University of Adelaide}
  \country{Australia}
}

\renewcommand{\shortauthors}{M. Crawford, W. Wang, R. Sun, and M. Xue}
 \renewcommand \authors{Matthew Crawford, Wei Wang, Ruoxi Sun, and Minhui Xue}

\begin{abstract}
With the rapid growth of malware attacks, more antivirus developers consider deploying machine learning technologies into their productions. Researchers and developers published various machine learning-based detectors with high precision on malware detection in recent years. Although numerous machine learning-based malware detectors are available, they face various machine learning-targeted attacks, including evasion and adversarial attacks. This project explores how and why adversarial examples evade malware detectors, then proposes a randomised chaining method to defend against adversarial malware statically. This research is crucial for working towards combating the pertinent malware cybercrime.
\end{abstract}

\maketitle

\section{Introduction}
Cybercrime is becoming more common in today's virtual ecosystem due to the global shift to digitising everything, such as the digital contact tracing during the pandemic, the emerging of IoT devices, and the tremendous spread of mobile applications~\cite{sun2021empirical,Feng2021SnipuzzBF,sun2020quality,sun2020venuetrace,sun2021understanding}. Cybercriminals use sophisticatedly disguised malware to evade detection from antivirus software while attacking systems and networks. Recent research~\cite{recent_research} has shown that the total malware infection growth rate has significantly increased during the past decade (2009 - 2018). In 2018, 812.67 million users were infected by malware, while malware infections were 12.4 million in 2009. It indicates that traditional rule-based malware detection products are far not enough to defend rapidly growing malware attacks. Meanwhile, machine learning technology has been adopted in commercial antivirus products. For instance, two world-leading antivirus engines, AVAST and Kaspersky, claimed that they leverage machine learning detection technologies in their products~\cite{avast_ml, kasp_ml}. 

Machine learning has been a hot topic in the research of malware detection. Researchers train a machine-learning model with a set of malicious and benign samples, as a training set, and use another set of samples, as a testing set, to validate the accuracy of the machine-learning model identifying malware from the testing set. However, machine learning-based malware detection is not the ultimate answer. The accuracy of state-of-the-art machine learning-based detectors are significantly high, but they rely on existing samples and are vulnerable against sophisticated adversarial examples~\cite{li2021leverage,wen2021great,li2021hidden,xu2021explainability,li2020invisible,chan2021breaking}. Recent research shows that both machine learning-based and rule-based malware detectors have weaknesses in malware detection~\cite{wang2021exposing}. By adopting explainable feature-space perturbation and problem-space obfuscation, malware can effectively evade the detection by machine learning-based and rule-based detectors, which can be regarded as black-box. 

%Recent research has shown that machine learning techniques can be used to hide malware from most current static antivirus systems. Attacks are shown to be effective when using black-box models, and even more effective when directly optimising adversarial malware against a specific antivirus system. 

This project will focus on 5 main goals, \one~research current techniques for optimising adversarial examples; \two~implement techniques and compare effectiveness of each; \three~explore how to detect the optimised adversarial malware; \four~propose a method to develop next-generation antivirus software; and \five~conduct an experiment to determine validity of the method. By the end of this project we will have obtained an extensive understanding of adversarial attacks and how next-generation antivirus detectors should operate.

\section{Background and related work}
To better understand the weaknesses of detectors defending malware, we conduct a background study on the state-of-the-art adversarial sample generation and evasion attacks.

\vspace{1mm}
\noindent \textbf{Functionality-preserving black-box optimization of adversarial windows malware.~}
Demetrio~\etal~\cite{demetrio2021functionality} implemented black box machine learning techniques to optimise adversarial Windows malware. The attack framework used to optimise the adversarial samples was called GAMMA (Genetic Adversarial Machine learning Malware Attack). It aimed to optimise the trade-off between payload size and misclassification confidence. The paper showed these attacks up against two static machine learning antivirus detectors called MalConv, a convolutional neural network, and GBDT (Gradient boosting decision trees), which used a fixed representation of 2,381 features.

They were able to show that malware detectors are vulnerable to adversarial samples, even with only black-box query access. This new technique was able to be more \one~query-efficient by injecting specific content for evasion purposes and \two~functionality preserving, where benign content, that will never be executed, is injected at the end of the file, or in newly created sections. This attack framework is ineffective against dynamic detectors since the functionality of the malware, before and after optimisation, remains the same.

This paper was able to show how effective black-box optimisations can be against static detectors, it is likely that with a few more papers, the adversarial malware may become near undetectable to static antivirus detectors. This should lead the security industry to start thinking about producing dynamic machine learning detectors that are fast and efficient. An interesting area to proceed would be to create countermeasures for dynamic malware detectors. One way could be to inject benign content into the malware that executes at runtime to throw off the detectors.

\vspace{1mm}
\noindent \textbf{Reinforcement learning attacks.~}
Song~\etal~\cite{song2020mab} implemented black box machine learning techniques to optimise adversarial Windows malware. The attack framework used was a multi-armed bandit (MAB) which had three main focuses, \one~limit exploration space by modelling the generation process as a stateless process, \two~reuse successful payloads in modelling and \three~minimise the changes on adversarial examples to correctly assign rewards. The two main components used to achieve this was the binary rewriter and the content minimiser. Essentially, the binary rewriter would generate a set of evasive samples that have been made to barely evade a target classifier. The content minimiser would then experiment by removing macro actions, then also micro actions, that would not effect the classification of the adversarial examples. This allowed rewards to be more  correctly assigned to the features that impacted evasiveness. 

This attack framework produced adversarial examples with high functionality preservation rates (binary remained malicious) and high functionality rates (binary would run 96\% of the time). Reducing the amount of failed attempts makes it more efficient to produce adversarial malware. The minimisation step allowed for more effective training of the model by first narrowing down the various optimisations while keeping the binary classified as malicious, then assigning rewards only to the necessary optimisations. The transferability rate between machine learning malware classifiers was considered “high” with a rating of 10\% from MalConv to EMBER and 23\% from EMBER to MalConv. The transferability rate between non-machine learning and machine learning detectors was less than 5\% for all cases. This means a new model should be trained for each new antivirus detector. The attack framework didn't apply any optimisations that would evade against dynamic detectors, authors considering this as out of scope.

This paper shows an efficient approach to training their model which greatly improves the performance of their adversarial examples against static detectors. Perhaps more macro and micro optimisations could be implemented in the attempt to increase performance. Although evading dynamic detectors was not considered in scope for this paper, perhaps a similar minimisation methodology could be applied to dynamic optimisations in some way.

\vspace{1mm}
\noindent \textbf{Explainability-guided evasion attacks.~}
Wang~\etal~\cite{wang2021exposing} evaluate the weaknesses of malware detectors with explainability-guided evasion attacks. They perform evasion attacks by generating adversarial samples from Android (APK), Windows (WinPE) and Linux (ELF) binaries. The adversarial sample generation methodologies include: \one~feature-space manipulation (FSM) by perturbing features to make it appear more benign; and \two~problem-space obfuscation (PSO) by modifying the code to make it less obvious what it is trying to achieve. The authors conduct FSM on three machine learning models: support vector machines (SVM), lightGBM and a simple-structured neural network. PSO leverages four obfuscation techniques: control-flow graph alternation, dead-code insertion, instruction substitution, and encryption. 

The result shows that attacks are effective against both non-learning and learning-based detectors. FSM and PSO are used to enlarge the attack surface and allow attacks to be performed in a black-box manner. The paper then demonstrates the attacks being used with three different types of binaries. It also explains the transferability of models, stating that the transferability of evasion attacks depends on the overlaps of decisive features among different learning-based models. Problem space obfuscation was less effective in regards to WinPE and ELF binaries since only the encryption strategy could be applied to them. Dynamic antivirus detectors are still effective against the presented attack strategies.

The explainability-guided approach was useful for conveying the strategies that were used in a way that is logical. The results were displayed in an easy-to-follow manner with key takeaways highlighted. It would be interesting to see this approach applied for evasion attacks against dynamic detectors, answering what works, does not work and why.

\vspace{1mm}
\noindent \textbf{Automatically evading classifiers.~}
Xu~\cite{xu2016automatically} proposes a generic and automatic methodology to generate PDF malware variants and evade the detection by malware classifiers. The methodology uses genetic programming (GP) techniques, a type of evolutionary algorithm, to generate evasive samples. It leverages prediction scores made by machine learning classifiers and rough knowledge of features used by the classifiers. In the experiment, the authors evaluated two PDF malware classifiers: PDFrate~\cite{pdfrate} and Hidost~\cite{hidost}. 

By analyzing the efficacy of mutation traces and variants on the classifiers, the authors found that the proposed methodology had a significantly high evasion rate when attacking both classifiers. In cross-evasion effects evaluation, generated variants still have a significant evasion rate against different classifiers.
Although malware variants generated by the proposed methodology shows a significant evasion rate on two PDF malware classifiers, we are concerned about the capability on other datasets, \eg Windows binaries. Another concern is that the high cross-evasion rate is based on limited classifiers that have overfitting issues. The authors should evaluate more detectors to evaluate the effectiveness and genericity of their methodology.

Leveraging GP techniques to generate malware variants seems evasive on PDFrate and Hidost. However, we should note that it is computationally expensive. The authors continuously executed their methodology for approximately one week for PDFrate and two days for Hidost to find effective mutation traces and generate variants. Therefore, purely using GP does not satisfy the requirement of large-scale empirical study and defends the fast evolution of malware. 

\section{Methodology}
In this section we introduce the methodology of our research. We start from the existing malware adversarial generating technologies, comparing the effectiveness of each approach, then exploring how to detect adversarial malware, and proposing a novel method to develop next-generation antivirus engine.
\subsection{Research current techniques for generating adversarial malware}
\vspace{1mm}
\noindent \textbf{MAB-Malware.~}
Song~\fancyref{song2020mab} produced this framework that trains a reinforcement learning model. It consists of two main modules, the binary rewriter and the action minimiser. It iterates over the binary rewriter and action minimiser modules until an evasive sample is generated or it exceeds the total number of max attempts. The main improvements over other frameworks is that it \one~models the generation as a stateless process, \two~reuses successful payloads and \three~minimises changes to correctly assign rewards.

The binary rewriter uses Thompson sampling which operates by maximising some expected reward with respect to some some randomly drawn belief to select an action sequence. It then rewrites it to gain variants, and if a variant then evades the target detector, it is sent on to the action minimiser. The two types of actions that can be applied are macro and micro, where micro is only injecting 1 byte, whereas macro is more than that. A machine has a set of actions assigned to it and follows a beta distribution specific to a tuple denoted by $\alpha$ and $\beta$. For each sample generated from the machine, for each action used, when evasive, $\alpha$ is increased by 1, else $\beta$ is increased by 1. After a few trials $\alpha$ and $\beta$ become large and the uncertainty decreases. Because of this, machines with low values are used for exploration and machines with high average rewards are selected for exploitation.

The action minimiser iteratively removes actions then tests evasiveness of the adversarial sample. It starts by removing macro actions while keeping the sample classified as benign, then micro actions while also keeping it classified as benign. Rewards are finally applied to the actions that were required for a benign classification. This process is able to produced a minimised evasive sample.

\vspace{1mm}
\noindent \textbf{GAMMA.~}
Demetrio~\fancyref{demetrio2021functionality} produced another black box attack framework that relies upon a set of functionality preserving manipulations that inject content into the malicious program without altering its execution traces. Content is extracted from benign samples rather than being randomly generated. It is formulated as a constrained optimisation problem that minimises both probability to evade detection and size of injected content. The goal being to inject the minimum amount of content to become evasive while maintaining the files malicious properties. The main components of GAMMA are \one~payload generation, \two~payload injection and \three~evaluation.

The solution algorithm is a genetic algorithm which iterates over three steps, selection, cross-over and mutation. In the selection phase, the objective function is used to evaluate candidates and choose a sample of the best candidates. The cross-over phase takes the best candidates and randomly mixes around values with other candidates in the selection. Finally, the mutation phase further randomises candidate vectors by changing elements of each input vector at random with a low probability. The newly generated candidate vectors get unioned with the initial vectors then the cycle repeats, starting from the selection step. The amount of iterations has a predefined upper bound. After it is reached, the candidate which has the lowest objective function value is returned. 

\subsection{Implement techniques and compare effectiveness of each}

\vspace{1mm}
\noindent \textbf{Testing MAB framework.~}
Two tests were ran using the MAB Malware framework. The first was a collection of 1000 malware samples obtained from the default docker image available at the MAB-Malware GitHub repository\footnote{https://github.com/weisong-ucr/MAB-malware}, which was optimised against an EMBER detector. The second was a collection of 2448 malware samples obtained from the University of Adelaide for research purposes, which was optimised against a MalConv Detector. 

The first test had a runtime of approximately 4 hours, whereas the second was approximately 6 hours and both produced 810 minimised evasive samples before completing. The transferability from the first evasive samples to a current EMBER model was 47\%, whereas the second evasive samples to an EMBER model was 28\%. The final machines show that overlay append was the most successful action for producing evasive adversarial samples against EMBER and MalConv.

\vspace{1mm}
\noindent \textbf{Testing GAMMA framework.~}
GAMMA tests required both benign and malicious files to produce adversarial samples. A collection of 2448 malware samples obtained from the University of Adelaide and 1182 benign samples from files found in the Windows operating system were gathered for the tests. All attacks were ran against a pre-trained MalConv model then the transferability was further analysed by running it against an EMBER model. 

All attacks were supplied the 1112 benign files, the independent variable being the amount of malicious files. The first attack used 1 malicious file, had a 16 second run time and was not evasive to either detector. The second and third attacks both used 10 malicious files, had an average of 2 minute run time and was both able to evade the MalConv detector. However, both adversarial samples were unable to evade EMBER. The final test used 50 malicious samples, had a run time of 10 minutes and was able to evade MalConv. The adversarial sample was unable to evade EMBER, however, it was given a more benign rating of 0.75 compared to the previous lowest being 0.9. 

\vspace{1mm}
\noindent \textbf{Compare the frameworks.~}
MAB and GAMMA are both frameworks that optimise malware against antivirus detectors by using black-box attacks. The output for MAB is significantly more detailed than GAMMA which allows larger tests to be ran because even if it fails, there is plenty of information to understand what the program was doing. MAB was also more reliable when it came to producing working malicious binaries. For these reasons, the MAB framework will be used later when exploring ideas on how to defend against adversarial attacks.

\subsection{Explore how to detect adversarial malware}
As discussed previously, the transferability of adversarial malware is currently low between different antivirus detectors. Findings from Song~\fancyref{song2020mab} support these findings and also highlight how transferability is generally higher between detectors that consider similar features as important. Exploiting this common weakness in adversarial malware attacks could prove an effective way to also defend against them.

Since transferability is low, perhaps chaining together multiple detectors that focus on different aspects for detecting malware could make a system that is more difficult to perform adversarial attacks against. In this case, chaining would involve starting with a fixed set of detectors then have them scan a file until one of them returns a malicious rating. The file is considered benign only if all detectors rate it so. Ideally, the detectors should be uniquely effective in at least a single way to maximise detection rate. Some weaknesses/limitations of this chaining method include a slower runtime, the assumption that all detectors are effective in some way and increased chance of false positives. However, the biggest weakness of this method is that it does not guarantee that adversarial attacks can not just be targeted against the chain as a whole. With enough computations, it is likely evasive adversarial samples will eventually be produced. As long as the attacker can predict the software defending the machine at the time of the attack, it is unlikely to formulate an effective static defence.

\subsection{Propose a method to develop next-generation antivirus software}
If an attacker can predict the software protecting a system, then a successful adversarial attack is theoretically possible. Improving upon the previously mention chaining method, a randomised chaining method would make the antivirus software in the chain near impossible to predict. For this system, a large collection of malware detectors is optimal. When a new scan queue is created, a defined amount of detectors will be randomly selected from the collection. The selected detectors would then classify the queued files using the previously mention chaining method.

The main advantage of this method is that it removes the ability for an attacker to predict the antivirus software in use. It also has the ability to draw from a collection so large that it would be unreasonable for an attacker to attempt optimisations against all detectors within, while still having the run time of chaining a small number of detectors. The randomness aspect of this method allows the average detection effectiveness to be reached with just a sample of the total collection. In the next section an experiment will be conducted to determine theoretical mean detection rate using different sample sizes.
This method carries across limitations from the chaining method since it never addressed them. These limitations include a slow runtime, increased chance of false positives and the assumption that each detector is effective in some way. This system also requires a large amount detectors to maximise detection rate and increase protection to adversarial attacks. For future work, it may be possible to generate many detectors using different features from a wide range of malware samples.

\section{Experiment and Result}
%In this paper, we use 200 adversarial samples that were randomly selected from generated, evasive outputs as a result of using the MAB framework on two datasets. The first was a collection of 1000 malware samples obtained from using the default docker image available from the MAB-Malware GitHub repository\footnote{https://github.com/weisong-ucr/MAB-malware}, which was optimised against an EMBER detector. The second was a collection of 2448 malware samples obtained from the University of Adelaide for research purposes, which was optimised against a MalConv Detector. 100 samples were then collected from each evasive output to formulate our dataset.

In this section, we deploy the MAB-Malware source code~\footnote{https://github.com/weisong-ucr/MAB-malware} and generate adversarial samples from two datasets. The first dataset is obtained from the MAB-Malware project which contains 1000 malware samples. These samples are Windows Pocket Executive binaries and optimized against Ember~\cite{anderson2018ember}, a LightGBM model. The second dataset is a collection of 2448 malware samples obtained from the University of Adelaide for research purposes, which was optimised against a MalConv Detector. Then we randomly choose 200 generated adversarial samples from the previous step.

The quality and validity of the dataset was not verified, instead it is assumed that, through random selection and using a medium sample size, it is likely that the data is adequate for this experiment. Improving upon this is left as future work.

\begin{figure}[htp]
    \centering
    \includegraphics[width=\linewidth]{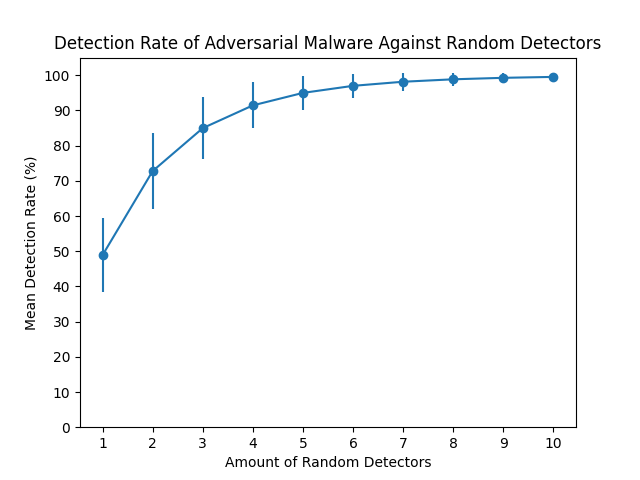}
    \caption{A higher mean detection rate is obtained when the amount of random detectors increases. Vertical lines show standard deviation at different points.}
    \label{result}
\end{figure}

For the experiment, all 200 of the selected adversarial samples were submitted to Virus Total which were then scanned by 67 antivirus detectors. The detection rate (malicious ratings / total detectors) of each adversarial sample was then saved into a csv file. A script was then used to graph the theoretical effectiveness of the randomised chaining method. The script calculated the mean average and standard deviation of the detection rates after using different amounts of random detectors. Figure \ref{result} shows the mean detection rate against amount of random detection with the standard deviation represented as a small line on each increment.

Figure \ref{result} shows that as the amount of random detectors increase, the detection rate of the randomised chaining method also increases. The average detection rate approaches 100\% as the amount of detectors increase. With just 5 random detectors, there is a 95\% chance that an adversarial sample is detected with a standard deviation of 4.8\%. Using 10 random detectors, there is a 99.5\% detection rate with a standard deviation of 1\%. 

These results show that targeting the low transferability of adversarial malware can be a viable strategy to statically defending against it. However, figure \ref{result} only shows the theoretical results of if we were to build the system and run attacks against it using the antivirus detectors found on Virus Total. In reality, results may vary depending on the amount and quality of antivirus software in the collection. All of the adversarial samples submitted to Virus Total to collect raw data were generated by running an attack against both MalConv and EMBER detectors separately. This does not produce a good representation of attacking the system since neither of those detectors are used as part of the Virus Total collection. However, Virus Total includes a wide range on antivirus solutions which is optimal for the randomised chaining method. Finally, it is important to note that the MAB framework does not aim to be highly transferable between many antivirus systems, instead it would retrain against any new system it encounters. Therefore, an adversarial attack that focuses on high transferability could prove effective against this system and significantly alter the experimental results. 
% Code for producing these results can be found in the project GitHub repository\footnote{https://github.cs.adelaide.edu.au/a1796164/TCS\_project\_2021\_s2.git}.

\section{Conclusion}
We discover that adversarial attacks produce malware with low transferability between detectors. We then design a randomised chaining method that could be effective at statically defending against current adversarial attacks because of this weakness. It operates by having a diverse collection of antivirus detectors where subsets of a predetermined size will be randomly selected from. The detectors in the subset will all scan a file and will classify that file as malicious if any single detector deems it so. Our results show that with a subset size of 10, there is a 99.5\% chance of detecting an adversarial sample with 1\% standard deviation. 

% \newpage
% \bibliographystyle{ACM-Reference-Format}
\balance
\bibliographystyle{unsrt}
\bibliography{references}

\begin{thebibliography}{10}

\bibitem{sun2021empirical}
Ruoxi Sun, Wei Wang, Minhui Xue, Gareth Tyson, Seyit Camtepe, and Damith~C
  Ranasinghe.
\newblock An empirical assessment of global covid-19 contact tracing
  applications.
\newblock In {\em 2021 IEEE/ACM 43rd International Conference on Software
  Engineering (ICSE)}, pages 1085--1097. IEEE, 2021.

\bibitem{Feng2021SnipuzzBF}
Xiaotao Feng, Ruoxi Sun, Xiaogang Zhu, Minghui Xue, Sheng Wen, Dongxi Liu,
  Surya Nepal, and Yang Xiang.
\newblock Snipuzz: Black-box fuzzing of iot firmware via message snippet
  inference.
\newblock {\em Proceedings of the 2021 ACM SIGSAC Conference on Computer and
  Communications Security}, 2021.

\bibitem{sun2020quality}
Ruoxi Sun and Minhui Xue.
\newblock Quality assessment of online automated privacy policy generators: an
  empirical study.
\newblock In {\em Proceedings of the Evaluation and Assessment in Software
  Engineering}, pages 270--275. 2020.

\bibitem{sun2020venuetrace}
Ruoxi Sun, Wei Wang, Minhui Xue, Gareth Tyson, and Damith~C Ranasinghe.
\newblock Venuetrace: a privacy-by-design covid-19 digital contact tracing
  solution.
\newblock In {\em Proceedings of the 18th Conference on Embedded Networked
  Sensor Systems}, pages 790--791, 2020.

\bibitem{sun2021understanding}
Suibin Sun, Le~Yu, Xiaokuan Zhang, Minhui Xue, Ren Zhou, Haojin Zhu, Shuang
  Hao, and Xiaodong Lin.
\newblock Understanding and detecting mobile ad fraud through the lens of
  invalid traffic.
\newblock In {\em Proceedings of the 2021 ACM SIGSAC Conference on Computer and
  Communications Security}, 2021.

\bibitem{recent_research}
Cyber security statistics.

\bibitem{avast_ml}
Training the avast machine learning engine.

\bibitem{kasp_ml}
Machine learning in cybersecurity.

\bibitem{wen2021great}
Jialin Wen, Benjamin Zi~Hao Zhao, Minhui Xue, Alina Oprea, and Haifeng Qian.
\newblock With great dispersion comes greater resilience: Efficient poisoning
  attacks and defenses for linear regression models.
\newblock {\em IEEE Transactions on Information Forensics and Security}, 2021.

\bibitem{li2021hidden}
Shaofeng Li, Hui Liu, Tian Dong, Benjamin Zi~Hao Zhao, Minhui Xue, Haojin Zhu,
  and Jialiang Lu.
\newblock Hidden backdoors in human-centric language models.
\newblock {\em Proceedings of the 2021 ACM SIGSAC Conference on Computer and
  Communications Security}, 2021.

\bibitem{xu2021explainability}
Jing Xu, Minhui Xue, and Stjepan Picek.
\newblock Explainability-based backdoor attacks against graph neural networks.
\newblock In {\em Proceedings of the 3rd ACM Workshop on Wireless Security and
  Machine Learning}, 2021.

\bibitem{li2020invisible}
Shaofeng Li, Minhui Xue, Benjamin Zhao, Haojin Zhu, and Xinpeng Zhang.
\newblock Invisible backdoor attacks on deep neural networks via steganography
  and regularization.
\newblock {\em IEEE Transactions on Dependable and Secure Computing}, 2020.

\bibitem{chan2021breaking}
Alvin Chan, Lei Ma, Felix Juefei-Xu, Yew-Soon Ong, Xiaofei Xie, Minhui Xue, and
  Yang Liu.
\newblock Breaking neural reasoning architectures with metamorphic
  relation-based adversarial examples.
\newblock {\em IEEE Transactions on Neural Networks and Learning Systems},
  2021.

\bibitem{wang2021exposing}
Wei Wang, Ruoxi Sun, Tian Dong, Shaofeng Li, Minhui Xue, Gareth Tyson, and
  Haojin Zhu.
\newblock Exposing weaknesses of malware detectors with explainability-guided
  evasion attacks.
\newblock {\em arXiv preprint arXiv:2111.10085}, 2021.

\bibitem{demetrio2021functionality}
Luca Demetrio, Battista Biggio, Giovanni Lagorio, Fabio Roli, and Alessandro
  Armando.
\newblock Functionality-preserving black-box optimization of adversarial
  windows malware.
\newblock {\em IEEE Transactions on Information Forensics and Security},
  16:3469--3478, 2021.

\bibitem{song2020mab}
Wei Song, Xuezixiang Li, Sadia Afroz, Deepali Garg, Dmitry Kuznetsov, and Heng
  Yin.
\newblock Mab-malware: A reinforcement learning framework for attacking static
  malware classifiers.
\newblock {\em arXiv preprint arXiv:2003.03100}, 2020.

\bibitem{xu2016automatically}
Weilin Xu, Yanjun Qi, and David Evans.
\newblock Automatically evading classifiers.
\newblock In {\em Proceedings of the 2016 network and distributed systems
  symposium}, volume~10, 2016.

\bibitem{pdfrate}
Charles Smutz and Angelos Stavrou.
\newblock Malicious pdf detection using metadata and structural features.
\newblock In {\em Proceedings of the 28th Annual Computer Security Applications
  Conference}, ACSAC '12, 2012.

\bibitem{hidost}
Pavel Laskov.
\newblock Detection of malicious pdf files based on hierarchical document
  structure.
\newblock In {\em In Proceedings of the Network and Distributed System Security
  Symposium, NDSS 2013}. Citeseer, 2013.

\bibitem{anderson2018ember}
Hyrum~S Anderson and Phil Roth.
\newblock Ember: An open dataset for training static {PE} malware machine
  learning models.
\newblock {\em arXiv preprint arXiv:1804.04637}, 2018.

\end{thebibliography}

\end{document}